\documentstyle[preprint,tighten,pra,aps,latexsym,amssymb]{revtex}

\newcommand{\eq}{\begin{equation}}
\newcommand{\feq}{\end{equation}}
\newcommand{\eqn}{\begin{eqnarray}}
\newcommand{\feqn}{\end{eqnarray}}
\newcommand{\arr}{\begin{eqnarray*}}
\newcommand{\farr}{\end{eqnarray*}}

\newcommand{\R}{{\mathbb R}}

\begin{document}
\tightenlines
\draft

\def\al{\alpha}
\def\be{\beta}
\def\ga{\gamma}
\def\de{\delta}
\def\ep{\varepsilon}
\def\ze{\zeta}
\def\io{\iota}
\def\ka{\kappa}
\def\la{\lambda}
\def\roh{\varrho}
\def\si{\sigma}
\def\om{\omega}
\def\ph{\varphi}
\def\th{\theta}
\def\te{\vartheta}
\def\up{\upsilon}
\def\Ga{\Gamma}
\def\De{\Delta}
\def\La{\Lambda}
\def\Si{\Sigma}
\def\Om{\Omega}
\def\Te{\Theta}
\def\Th{\Theta}
\def\Up{\Upsilon}

\preprint{UTF 426}

\title{Rotating Black Branes wrapped on Einstein Spaces}

\author{Dietmar Klemm\footnote{email: klemm@science.unitn.it}\\ 
\vspace*{0.5cm}}

\address{Universit\`a  degli Studi di Trento,\\
Dipartimento di Fisica,\\
Via Sommarive, 14\\
38050 Povo (TN)\\
Italia\\
\vspace*{0.5cm}      
and\\ Istituto Nazionale di Fisica Nucleare,\\
Gruppo Collegato di Trento,\\ Italia}

\maketitle
\begin{abstract}
We present new rotating black brane solutions which solve
Einstein's equations with cosmological constant $\Lambda$
in arbitrary dimension $d$.
For negative $\Lambda$, the branes naturally appear in
AdS supergravity compactifications, and should therefore play some
role in the AdS/CFT correspondence. The spacetimes are warped products
of a four-dimensional part and an Einstein space of dimension $d-4$,
which is not necessarily of constant curvature.
As a special subcase, the solutions contain
the higher dimensional generalization of the Kerr-AdS metric recently
found by Hawking et al.
\end{abstract}


\maketitle

\section{Introduction and Motivation}
The AdS/CFT correspondence \cite{malda,kleb,witten1,witten2} relates
conformal field theory on a manifold $\cal B$ to 
supergravity or string theory on
${\cal M} \times {\cal K}$, where $\cal K$ is a compact space,
and $\cal M$ denotes an Einstein manifold with negative cosmological
constant, which is required to possess a conformal compactification
$\bar{\cal M}$ with boundary $\partial\bar{\cal M} = {\cal B}$.
The conformal field theory partition function, $Z_{CFT}({\cal B})$, is then
conjectured to be a sum over all ${\cal M}_i$ with
$\partial\bar{{\cal M}_i} = {\cal B}$,
\eq
Z_{CFT}({\cal B}) = \sum_i Z_S({\cal M}_i), \label{corr}
\feq
where $Z_S({\cal M}_i)$ is the string theory partition function on
${\cal M}_i$.\\
Given this correspondence, it is interesting to look for new bulk
supergravity spacetimes representing Einstein spaces with negative
cosmological constant, as they furnish important information on
the conformal field theory on their boundary by means of (\ref{corr}),
e.~g.~information on phase transitions etc.\\
Of particular interest in this context is the relationship between
rotating black objects in the bulk and the conformal field theory
on the boundary. This has been examined recently in \cite{taylor} for
Kerr-AdS black holes. (See also \cite{csaki}, where rotating D-branes
have been used to study large $N$ QCD).
Now already in four dimensions there is a
variety of black configurations apart from the Kerr-AdS solution.
A class of such objects was presented in \cite{kmv} as stationary
generalizations of so-called topological black holes. In the present
paper we shall find higher-dimensional versions of these solutions,
generalizing the $d$-dimensional Kerr-AdS metric recently found in
\cite{taylor} to arbitrary topologies. The found solutions represent,
in the broadest sense,
rotating black branes which can wrap around Einstein spaces. They
are different from the spinning M-branes studied previously in the
literature \cite{cvetic,townsend}. We expect them to
provide an interesting ground to study the AdS/CFT correspondence.

\section{The Black Brane Solution} 
The $d$-dimensional Kerr-AdS metric presented in \cite{taylor},
is given by
\eqn
ds^2 & = & - \frac{\Delta_r}{\rho^2}[dt - \frac{a}{\Xi}
           \sin^2\theta d\phi]^2 + 
           \frac{\rho^2}{\Delta_r}dr^2 + \frac{\rho^2}{\Delta_\theta}d\theta^2
           \nonumber \\
       & & + \frac{\Delta_\theta\sin^2\theta}{\rho^2}[adt -
           \frac{(r^2+a^2)}{\Xi} d\phi]^2
          + r^2\cos^2\theta d\Omega_{d-4}^2, \label{dKerrAdS}
\feqn
where $d\Omega_{d-4}^2$ denotes the standard metric on the $(d-4)$-sphere
(later we shall see that this is actually too restrictive, because also
arbitrary Einstein metrics with positive scalar curvature are allowed for
$d\Omega_{d-4}^2$), and 
\eqn
\Delta_r & = & (r^2+a^2)(1+r^2l^{-2}) - 2\eta r^{5-d}, \nonumber \\
\Delta_\theta & = & 1 - a^2 l^{-2} \cos^2\theta, \\
\Xi & = & 1 - a^2 l^{-2}, \nonumber \\ 
\rho^2 & = & r^2 + a^2\cos^2\theta, \nonumber 
\feqn
where $\eta$ denotes the mass parameter.
In $d$ dimensions the rotation group is SO($d-1$), which has $[(d-1)/2]$
Casimir operators ($[x]$ denotes the integer part of $x$). Hence one
expects that the general spinning black hole solution is characterized
by $[(d-1)/2]$ angular momentum invariants \cite{perry}.
For $d=5$, a generalization
of (\ref{dKerrAdS}) to two rotation parameters was given in \cite{taylor},
but here we shall only consider the case of one angular momentum invariant.\\
The first thing we note is that we can analytically continue (\ref{dKerrAdS})
in order to obtain hyperbolical rather than spherical solutions.
To this aim, we write the metric on the $(d-4)$-sphere as
\eq
d\Omega_{d-4}^2 = d\chi^2 + \sin^2\chi d\Omega_{d-5}^2,
\feq
and substitute
\eqn
&& t \to it, \qquad r \to ir, \qquad \theta \to i\theta, \qquad \phi \to \phi,
\qquad \chi \to i\chi, \nonumber \\
&& a \to ia, \qquad \eta \to \eta/i^{5-d}
\feqn
in (\ref{dKerrAdS}), yielding
\eqn
ds^2 & = & - \frac{\Delta_r}{\rho^2}[dt + \frac{a}{\Xi}
           \sinh^2\theta d\phi]^2 +
           \frac{\rho^2}{\Delta_r}dr^2 + \frac{\rho^2}{\Delta_\theta}d\theta^2
           \nonumber \\
       & & + \frac{\Delta_\theta\sinh^2\theta}{\rho^2}[adt -
           \frac{(r^2+a^2)}{\Xi} d\phi]^2
          + r^2\cosh^2\theta d\Sigma_{d-4}^2, \label{dhyperbAdS}
\feqn
where now
\eqn
\Delta_r & = & (r^2+a^2)(-1+r^2l^{-2}) - 2\eta r^{5-d}, \nonumber \\
\Delta_\theta & = & 1 + a^2 l^{-2} \cosh^2\theta, \\
\Xi & = & 1 + a^2 l^{-2}, \nonumber \\ 
\rho^2 & = & r^2 + a^2\cosh^2\theta, \nonumber 
\feqn
and
\eq
d\Sigma_{d-4}^2 = d\chi^2 + \sinh^2\chi d\Omega_{d-5}^2
\feq
stands for the standard metric on hyperbolic $(d-4)$-space $H^{d-4}$. One
could also take a quotient $H^{d-4}/\mathrm{G}$, where $\mathrm{G}$
is a discrete subgroup of the isometry group SO($d-4,1$) of
$H^{d-4}$, acting properly discontinuously. E.~g.~for $d=6$, we can
have a Riemann surface of genus $g$. Of course, the same is valid
also in the spherical case, where also quotients of $S^{d-4}$ are
allowed, e.~g.~Lens spaces or projective spaces.\\
For $d=4$ the metric (\ref{dhyperbAdS}) reduces to the one found
in \cite{kmv} as a stationary generalization of four-dimensional
topological black holes. In this case a few comments are in order.
First of all, the statement in \cite{kmv}, that the $(\theta,\phi)$ sector
can be compactified to obtain Riemann surfaces of genus $g$ by identifying
opposite sites of a regular geodesic $4g$-gon centered at the origin
$(\theta,\phi) = (0,0)$, is not true. To see this, consider a
fundamental region chosen so
that the mid-point of one geodesic segment is at $(\theta,\phi) =
(\theta_0,0)$. The mid-point of the opposing segment is then at
$(\theta,\phi) = (\theta_0,\pi)$.  Let us denote these mid-points
respectively by $p_1$ and $p_2$. 
Consider the identification map $J$ that takes
the first segment to the second segment, and in particular $p_1$ to $p_2$. The
push-forward of $J$ then takes the set of one-forms $(dt,d\theta,d\phi)$
at $p_1$ to the set $(dt,-d\theta,-d\phi)$ at $p_2$.  As the coefficient of $dt
d\phi$ in the metric is the same at $p_1$ and $p_2$,
this means that $J$ does not
take the metric at $p_1$ to the metric at $p_2$,
so the metric would be discontinuous
\footnote{D.~K.~, like the other authors of \cite{kmv}, would like to
thank Jorma Louko for pointing out this.}.
Therefore, as it stands, the metric (\ref{dhyperbAdS}) for $d=4$
describes rotating black membranes in AdS space, and only in the
static case one can compactify them to obtain topological black holes.
For $d>4$, the same remains true for the $(\theta,\phi)$ sector of the
spacetime, however now we can compactify the additional space with
metric $d\Sigma^2_{d-4}$ as desribed above, without running into
difficulties.\\
Now the metrics (\ref{dKerrAdS}) and (\ref{dhyperbAdS}) are not
the end of the story. It is known \cite{kmv} (see also below)
that for $d=4$ they
emerge as a special case of the most general Petrov-type D metric
found by Plebanski and Demianski \cite{pleb}. This, in turn, contains
other black objects as subcases, e.~g.~a four-dimensional rotating cylindrical
AdS black hole \cite{kmv}. Besides, in view of the fact that
the horizon of $d$-dimensional static topological AdS black holes
is not necessarily a space of constant curvature, but can be an
arbitrary Einstein space \cite{birmingham}, we expect that we can replace the
constant curvature metrics $d\Omega^2_{d-4}$ and $d\Sigma^2_{d-4}$
used above by more general Einstein metrics.\\
The most general known Petrov-type D solution of Einstein's equations
with cosmological constant found in \cite{pleb} reads
\begin{eqnarray}
ds^2 &=& \frac{1}{(1 - pq)^2}\left\{\frac{p^2 + q^2}{\cal P}dp^2 +
         \frac{\cal P}{p^2 + q^2}(d\tau + q^2 d\sigma)^2\right. \nonumber \\ 
     & & \left.+ \frac{p^2 + q^2}{\cal L}dq^2 -
         \frac{\cal L}{p^2 + q^2}(d\tau - p^2 d\sigma)^2 \right\}, \label{PD}
\end{eqnarray}
where the structure functions are given by
\begin{eqnarray}
\cal P &=& \left(-\frac{\Lambda}{6} + \gamma\right) + 2np - \epsilon p^2
           + 2\eta p^3 + \left(-\frac{\Lambda}{6} - \gamma\right)p^4,
           \nonumber \\ 
\cal L &=& \left(-\frac{\Lambda}{6} + \gamma\right) - 2\eta q + \epsilon q^2
           - 2n q^3 + \left(-\frac{\Lambda}{6} - \gamma\right)q^4.
\end{eqnarray}
$\Lambda$ is the cosmological constant, $\eta$ and $n$ are the mass and nut
parameters, respectively, and $\epsilon$ and $\gamma$ are further
real parameters. (For details cf.~\cite{pleb}). Rescaling the coordinates
and the constants according to
\begin{eqnarray}
&&p \to L^{-1}p, \qquad q \to L^{-1}q, \qquad \tau \to L\tau, \qquad
  \sigma \to L^3\sigma,
  \nonumber \\ 
&&\eta \to L^{-3}\eta, \qquad \epsilon \to L^{-2}\epsilon, \qquad
  n \to L^{-3}n, \qquad
  \gamma \to L^{-4}\gamma + \frac{\Lambda}{6}, \qquad \Lambda \to \Lambda
\end{eqnarray}
and taking the limit as $L \to \infty$, one obtains
\begin{eqnarray}
ds^2 &=& \frac{p^2 + q^2}{\cal P}dp^2 +
         \frac{\cal P}{p^2 + q^2}(d\tau + q^2 d\sigma)^2 \nonumber \\ 
     & & + \frac{p^2 + q^2}{\cal L}dq^2 - \frac{\cal L}{p^2 + q^2}(d\tau -
         p^2 d\sigma)^2,
\end{eqnarray}
where now
\begin{eqnarray}
\cal P &=& \gamma + 2np - \epsilon p^2 - \frac{\Lambda}{3}p^4, \nonumber \\ 
\cal L &=& \gamma - 2\eta q + \epsilon q^2 - \frac{\Lambda}{3}q^4.
\end{eqnarray}
Setting now
\begin{eqnarray}
&&q = r, \qquad p = a\cosh\theta, \qquad \tau = t - \frac{a\phi}{\Xi},
  \qquad \sigma = 
  -\frac{\phi}{a\Xi}, \nonumber \\ 
&& \epsilon = -1 - \frac{\Lambda a^2}{3}, \qquad \gamma = -a^2, \qquad
  n = 0, \qquad \Lambda = -3l^{-2},
\end{eqnarray}
one gets our solution (\ref{dhyperbAdS}) for $d=4$, whereas for
\begin{eqnarray}
&&q = r, \qquad p = a\cos\theta, \qquad \tau = t - \frac{a\phi}{\Xi},
  \qquad \sigma = 
  -\frac{\phi}{a\Xi}, \nonumber \\ 
&& \epsilon = 1 - \frac{\Lambda a^2}{3}, \qquad \gamma = a^2, \qquad
  n = 0, \qquad \Lambda = -3l^{-2},
\end{eqnarray}
one obtains the four-dimensional Kerr-AdS metric.
In view of the structure of (\ref{dKerrAdS}) and (\ref{dhyperbAdS})
and the results obtained above (like $r^2\cos^2\theta \propto q^2p^2$),
we make the following warped product ansatz for the generalized metric
in $d$ dimensions
\eqn
ds^2 &=& \frac{p^2 + q^2}{\cal P}dp^2 +
         \frac{\cal P}{p^2 + q^2}(d\tau + q^2 d\sigma)^2 \nonumber \\ 
     & & + \frac{p^2 + q^2}{\cal L}dq^2 - \frac{\cal L}{p^2 + q^2}(d\tau -
         p^2 d\sigma)^2 + q^2p^2d\Omega_{d-4}^2, \label{genansatz}
\feqn
where we assume $d\Omega^2_{d-4}$ to be Einstein with scalar curvature
$(d-4)\kappa$, with $\kappa$ a constant. ${\cal P}(p)$ and ${\cal L}(q)$
are structure functions, which we have to determine. 
The Ricci tensor of
(\ref{genansatz}) can be computed by the following\\

\begin{it}\underline{Proposition}\cite{oneill}.
Let $M = B \times_f F$ be a warped
product, with $D = \mbox{dim}F>1$,
$X,Y$ horizontal (i.~e.~tangent to the basis $B$) and $V,W$
vertical (i.~e.~tangent to the fiber $F$) vector fields. Then\end{it}
\eqn
\mbox{Ric}(X,Y) &=& ^B\mbox{Ric}(X,Y) - (D/f)H^f(X,Y), \label{hor} \\
\mbox{Ric}(X,V) &=& 0, \\
\mbox{Ric}(V,W) &=& ^F\mbox{Ric}(X,Y) - g(V,W)f^{\#}, \label{vert}
\feqn
{\it where $g(.,.)$ denotes the metric on $M$,}
\eq
f^{\#} = \frac{\Delta f}{f} + (D-1)\frac{g(\mbox{grad}f,\mbox{grad}f)}{f^2},
\feq
{\it and $\Delta f$ and $H^f$ are the Laplacian and Hessian on $B$,
respectively.}\\

For the ansatz (\ref{genansatz}), we choose the following vierbein
on the basis $B$
\eqn
e^0 &=& \frac{\sqrt{\cal L}}{\rho}(d\tau - p^2d\sigma), \qquad
e^1 = \frac{\rho}{\sqrt{\cal L}}dq, \nonumber \\
e^2 &=& \frac{\rho}{\sqrt{\cal P}}dp, \qquad
e^3 = \frac{\sqrt{\cal P}}{\rho}(d\tau + q^2d\sigma),
\feqn
where we introduced
\eq
\rho^2 = p^2 + q^2.
\feq
The only nonvanishing
orthonormal frame components of the Ricci tensor $^B\mbox{Ric}$
of the basis manifold then read 
\eqn
^B\mbox{Ric}_{00} &=& -^B\mbox{Ric}_{11} = \frac{{\cal L}''\rho^2 + 2{\cal P}'p
                      - 2{\cal L}'q + 2{\cal L} - 2{\cal P}}{2\rho^4},
                      \nonumber \\
^B\mbox{Ric}_{22} &=& ^B\mbox{Ric}_{33} = -\frac{{\cal P}''\rho^2 - 2{\cal P}'p
                      + 2{\cal L}'q - 2{\cal L} + 2{\cal P}}{2\rho^4},
\feqn
where the primes denote derivatives with respect to $p$ or $q$.
For our warping function, $f = pq$, the Hessian is given by
\eqn
H^f_{00} &=& H^f_{11} = -\frac{({\cal L}'\rho^2 + 2q({\cal P}
                      - {\cal L}))p}{2\rho^4},
                      \nonumber \\
H^f_{22} &=& H^f_{33} = \frac{({\cal P}'\rho^2 - 2p({\cal P}
                      - {\cal L}))q}{2\rho^4},
\feqn
the off-diagonal components being zero.
Let us consider first eq.~(\ref{vert}). Requiring $\mbox{Ric}(V,W) =
\Lambda g(V,W)$ and $^F\mbox{Ric}(V,W) = \kappa g_F(V,W)$, where
$g_F(V,W) = g(V,W)/f^2$ is the fiber metric, and inserting the expressions
for the Laplacian and the gradient, one arrives at the equation
\eq
\Lambda p^2q^2\rho^2 = \kappa \rho^2 - qp({\cal P}'q + {\cal L}'p) -
                       (D-1)({\cal P}q^2 + {\cal L}p^2), \label{vert2}
\feq
with $D = d-4$ in our case. Applying the operator $\partial^4_p
\partial^2_q$ to (\ref{vert2}) yields the sixth order differential equation
\eq
(D+4){\cal P}^{(5)} + p{\cal P}^{(6)} = 0 \label{diffeq}
\feq
for ${\cal P}$, which has the solution
\eq
{\cal P} = C p^{1-D} + \sum_{i=0}^4\al_i p^i, \label{solP}
\feq
where $C \in \R$ and the $\al_i$ are integration constants.
Applying $\partial^2_p
\partial^4_q$ to (\ref{vert2}), we get the same differential equation
(\ref{diffeq}) for ${\cal L}$. This means
that we can write ${\cal L}$ in the form
\eq
{\cal L} = B q^{1-D} + \sum_{i=0}^4\be_i q^i, \label{solL}
\feq
$B, \be_i$ being integration constants. We now insert the results
(\ref{solP}) and (\ref{solL}) into (\ref{vert2}) and compare equal powers
of $p^iq^j$, which gives for the coefficients of the fourth-order
polynomials
\eqn
&& \kappa = (D-1)\al_0, \qquad \be_0 = \al_0, \qquad \al_1 = \be_1 = 0,
\qquad \al_2 = -\be_2, \nonumber \\
&& \al_3 = \be_3 = 0, \qquad \La = -(3+D)\al_4, \qquad \be_4 = \al_4.
\feqn
One finds that with these coefficients the horizontal equation
(\ref{hor}) is automatically satisfied, so we have found
the final solution
\eqn
{\cal P} &=& C p^{5-d} + \frac{\kappa}{d-5} + \al_2 p^2 -
             \frac{\La}{d-1}p^4, \nonumber \\
{\cal L} &=& B q^{5-d} + \frac{\kappa}{d-5} - \al_2 q^2 -
             \frac{\La}{d-1}q^4. \label{finalsol1}
\feqn
In the special case $d=5$ (corresponding to fiber dimension $D=1$),
the above proposition is not valid, so one has to carry out a separate
calculation, yielding
\eqn
{\cal P} &=& C + \al_2 p^2 - \frac{\La}{4}p^4, \nonumber \\
{\cal L} &=& B - \al_2 q^2 - \frac{\La}{4}q^4. \label{finalsol2}
\feqn
Starting from the general metric (\ref{genansatz}) with structure
functions given by (\ref{finalsol1}), we can recover the
$d$-dimensional Kerr-AdS solution (\ref{dKerrAdS}) by setting
\eqn
&& B = -2\eta, \qquad C = 0, \qquad \al_2 = -(1+a^2l^{-2}), \qquad
\kappa = a^2(d-5), \qquad \Lambda = -(d-1)l^{-2}, \nonumber \\
&&q = r, \qquad p = a\cos\theta, \qquad \tau = t - \frac{a\phi}{\Xi},
\qquad \sigma = -\frac{\phi}{a\Xi},
\feqn
with similar formulas for the hyperbolical case (\ref{dhyperbAdS}).
An interesting feature we notice is that the metric $d\Om^2_{d-4}$
occuring in (\ref{dKerrAdS})
is not necessarily of constant curvature; it is merely required to
be Einstein with positive scalar curvature $(d-5)(d-4)$.\\
A different solution can be obtained by the choice
\eqn
&& B = -2\eta, \qquad C = 0, \qquad \al_2 = 0, \qquad
\kappa = a^2(d-5), \qquad \Lambda = -(d-1)l^{-2}, \nonumber \\
&&q = r, \qquad p = a\zeta, \qquad \tau = t,
\qquad \sigma = \frac{\phi}{a},
\feqn
leading to
\eqn
ds^2 & = & - \frac{\Delta_r}{\rho^2}[dt - a
           \zeta^2 d\phi]^2 + 
           \frac{\rho^2}{\Delta_r}dr^2 + \frac{\rho^2}{\Delta_\zeta}d\zeta^2
           \nonumber \\
       & & + \frac{\Delta_\zeta}{\rho^2}[adt +
           r^2 d\phi]^2
          + r^2\zeta^2 d\Omega_{d-4}^2, \label{dcylAdS}
\feqn
where again $d\Omega_{d-4}^2$ denotes the metric of an Einstein space with
positive scalar curvature $(d-5)(d-4)$, and 
\eqn
\Delta_r & = & a^2 - 2\eta r^{5-d} +r^4l^{-2}, \nonumber \\
\Delta_\zeta & = & 1 + a^2 l^{-2} \zeta^4, \label{delzet}\\
\rho^2 & = & r^2 + a^2\zeta^2. \nonumber 
\feqn
(\ref{dcylAdS}) is a $d$-dimensional generalization of the
cylindrical black hole presented in \cite{kmv} (cylindrical, because
the $(\zeta,\phi)$ sector has the topology of a cylinder ($0 \le \phi
\le 2\pi$ with endpoints identified, and $\zeta \in \R$)).\\
We can also recast (\ref{genansatz}) in the canonical form
\eq
ds^2 = -N^2d\tau^2 + \frac{\rho^2}{{\cal L}}dq^2 + \frac{\rho^2}{{\cal P}}dp^2
       + \frac{\Sigma^2}{\rho^2}[d\sigma - \omega d\tau]^2
       + q^2p^2 d\Om_{d-4}^2, \label{canform}
\feq
where
\eq
\Sigma^2 = {\cal P}q^4 - {\cal L}p^4,
\feq
and the lapse function $N$ and the angular velocity $\omega$ are
given by
\eqn
N^2 &=& \frac{{\cal P}{\cal L}\rho^2}{\Sigma^2}, \nonumber \\
\omega &=& - \frac{{\cal P}q^2 + {\cal L}p^2}{\Sigma^2}.
\feqn
Let us now briefly discuss why the found solutions can be regarded
in a certain sense as branes. First of all we note that the
$(\theta,\phi)$ or $(\zeta,\phi)$ sectors of (\ref{dhyperbAdS})
respectively (\ref{dcylAdS}) are membrane-like\footnote{In the case of
(\ref{dcylAdS}) these membranes are wrapped on a circle parametrized
by $\phi$.}, even if the curvature
of these membranes is not constant.
The $d-4$ additional dimensions form an Einstein
space, as we saw above. If this Einstein space is compact, the
additional dimensions can wrap around it.\\
Now, given the new black configurations, one would like to
calculate the Euclidean action, as this is needed in the
AdS/CFT correspondence according to (\ref{corr}) (in the supergravity
approximation). Furthermore, one also needs the Euclidean action, if one wishes
to discuss the thermodynamics of the black objects. In this
context it would be of interest to examine whether a first law
of black hole mechanics is valid. The calculation of
the Euclidean action was done in \cite{taylor} for (\ref{dKerrAdS}),
i.~e.~for the $d$-dimensional Kerr-AdS black hole. For e.~g.~
(\ref{dhyperbAdS}), the thermodynamical
discussion is hindered by the fact
that the $(\theta,\phi)$ sector of the metric is noncompact, and one
cannot define quantities like mass or angular momentum {\it per unit
volume} like it is usually done for $p$-branes with Poincar\'{e}
invariance on the world volume, as these quantities would still depend
on $\theta$. Especially, for $d=4$ (and this should hold also
for arbitrary $d$), one obtains a mass density proportional to
$\sinh\theta$, whereas the angular momentum density goes with
$\sinh^3\theta$ \cite{kmv}. Therefore we do not expect that a first
law will be valid for the {\it densities} of the conserved
quantities. If one were able to calculate a finite Euclidean action,
these difficulties would have overcome, but to this aim it is
not sufficient to subtract a suitably chosen background in order to
cancel the divergences in $r$, because the divergences in $\theta$
remain. The calculation of the Euclidean action remains therefore an
open issue, which we shall leave for future investigations.\\
Nevertheless, it is still possible to extract some interesting thermodynamical
informations, as we shall see. Let us carry out the analytical
continuation $\tau = -i{\cal T}$ in (\ref{canform}), whereupon we
obtain the "quasi-Euclidean" section
\eq
ds^2 = N^2d{\cal T}^2 + \frac{\rho^2}{{\cal L}}dq^2 +
       \frac{\rho^2}{{\cal P}}dp^2
       + \frac{\Sigma^2}{\rho^2}[d\sigma + i\omega d{\cal T}]^2
       + q^2p^2 d\Om_{d-4}^2. \label{quasieucl}
\feq
This metric would be the starting point for a calculation of
the Euclidean action \cite{gibbhawk}. The temperature of the black
object can be found by the requirement that no conical singularities
appear in (\ref{quasieucl}). We assume to have an event horizon
on a zero $q_H$ of ${\cal L}(q)$, i.~e.~${\cal L}(q_H)=0$. Using the
expansion
\eq
{\cal L}(q) = {\cal L}'(q_H)(q - q_H)
\feq
near $q=q_H$, and defining the new coordinate
\eq
Q = 2\sqrt{\frac{q-q_H}{|{\cal L}'(q_H)|}},
\feq
(\ref{quasieucl}) can be written near $q=q_H$
\eq
ds^2 = \rho_H^2(dQ^2 + \frac{{\cal L}'(q_H)^2 Q^2}{4q_H^4}d{\cal T}^2)
       + \frac{\rho_H^2}{{\cal P}}dp^2
       + \frac{\Sigma_H^2}{\rho_H^2}[d\sigma + i\omega_H d{\cal T}]^2
       + q_H^2p^2 d\Om_{d-4}^2,
\feq
from which we see that the period of ${\cal T}$ must be
\eq
\beta = \frac{4\pi q_H^2}{|{\cal L}'(q_H)|}
\feq
in order to avoid conical singularities. (Besides, regularity of the
metric also requires to identify $\sigma \sim \sigma + i\beta \omega_H$
\cite{gibbhawk}, where $\omega_H = -1/q_H^2$ is the angular velocity on
the horizon). The Hawking temperature is then
\eq
T = \beta^{-1} = \frac{|{\cal L}'(q_H)|}{4\pi q_H^2}. \label{temp}
\feq
As an example, we calculate (\ref{temp}) for the generalized
cylindrical black hole spacetime (\ref{dcylAdS}), obtaining
\eq
T = \frac{r_H}{\pi l^2} - \frac{\eta(5-d)r_H^{2-d}}{2\pi}, \label{Tcyl}
\feq
where $r_H$ is a zero of $\Delta_r$ (\ref{delzet}), i.~e.~the
location of the event horizon. One observes that for $d>5$, the
temperature (\ref{Tcyl}) is always positive; it never becomes zero,
as $\eta>0$ in order to
have event horizons. We get a minimal temperature
\eq
T_{min} = \frac{r_H}{\pi l^2}\frac{d-1}{d-2}
\feq
for
\eq
r_H = \left(\frac{l^2\eta}{2}(d-5)(d-2)\right)^{\frac{1}{d-1}}.
\feq
This means that the black object contributes to the thermodynamics
of the corresponding boundary conformal field theory only if the
temperature is high enough, a behaviour similar to the
Hawking-Page phase transition \cite{hawkpage} from thermal AdS space
to the Schwarz\-schild-AdS black hole, which was studied in the
context of the AdS/CFT correspondence by Witten \cite{witten2}.
In the present case, the boundary conformal field theory would
live on $T^2 \times \R \times F$, where $T^2$ is a two-torus, and
$F$ an Einstein manifold of dimension $D>1$ with positive
scalar curvature, and due to the minimal temperature necessary
for the existence of the black hole in the bulk, we expect a phase
transition in this boundary conformal field theory.\\
The example above indicates that the found solutions may be interesting 
in the context of the AdS/CFT correspondence; a more precise study
of them could reveal further information on conformal field theories.

\section*{Acknowledgement}

This work has been supported by a research grant within the
scope of the {\em Common Special Academic Program III} of the
Federal Republic of Germany and its Federal States, mediated 
by the DAAD.


\newpage

\begin{thebibliography}{99}

\bibitem{malda} J.~M.~Maldacena, Adv.~Theor.~Math.~Phys.~{\bf 2}, 231 (1998).

\bibitem{kleb} S.~Gubser, I.~Klebanov, and A.~Polyakov, Phys.~Lett.~{\bf B428},
               105 (1998).

\bibitem{witten1} E.~Witten, Adv.~Theor.~Math.~Phys.~{\bf 2}, 253 (1998).

\bibitem{witten2} E.~Witten, Adv.~Theor.~Math.~Phys.~{\bf 2}, 505 (1998).

\bibitem{taylor} S.~W.~Hawking, C.~J.~Hunter, and M.~M.~Taylor-Robinson,
                 hep-th/9811056.

\bibitem{csaki} C.~Csaki, Y.~Oz, J.~Russo, and J.~Terning, hep-th/9810186.

\bibitem{kmv} D.~Klemm, V.~Moretti, and L.~Vanzo, Phys.~Rev.~D {\bf 57},
              6127 (1998).

\bibitem{cvetic} M.~Cvetic and D.~Youm, Nucl.~Phys.~{\bf B499}, 253 (1997).

\bibitem{townsend} J.~P.~Gauntlett, R.~C.~Myers, and P.~K.~Townsend,
                   hep-th/9809065.

\bibitem{perry} R.~C.~Myers and M.~Perry, Ann.~Phys.~(N.~Y.~) {\bf 172},
                304 (1986).

\bibitem{pleb} J.~F.~Plebanski and M.~Demianski, Ann.~Phys.~(N.~Y.~) {\bf 98},
                98 (1976).

\bibitem{birmingham} D.~Birmingham, hep-th/9808032.

\bibitem{oneill} B.~O'Neill, {\it Semi-Riemannian Geometry, with
                 Applications to Relativity}, Academic Press, San Diego, USA,
                 1983.

\bibitem{gibbhawk} G.~W.~Gibbons and S.~W.~Hawking, Phys.~Rev.~D {\bf 15},
                   2752 (1977).

\bibitem{hawkpage} S.~W.~Hawking and D.~Page, Commun.~Math.~Phys.~{\bf 87},
                   577 (1983).
  
\end{thebibliography}
\end{document}